\documentstyle[preprint,prl,aps,epsfig,psfig]{revtex}
\tightenlines
\newcommand{\be}{\begin{equation}}
\newcommand{\ee}{\end{equation}}
\newcommand{\bes}{\begin{eqnarray}}
\newcommand{\ees}{\end{eqnarray}}
\newcommand{\bfig}{\begin{figure}}
\newcommand{\efig}{\end{figure}}

\begin{document}
\title{Chaos or Noise --- Difficulties of a Distinction}
\author{ M. Cencini $^{(a,b)}$, M. Falcioni$^{(a)}$, 
 H. Kantz$^{(b)}$, E. Olbrich$^{(b)}$ and A. Vulpiani $^{(a)}$}
\address{$(a)$ Dipartimento di Fisica, Universit\'a di Roma ''La Sapienza''\\
 and INFM, Unit\'a di Roma, \\
p.le Aldo Moro 2, I-00185 Roma, Italy} 
\address{$(b)$ Max-Planck-Institut f\"ur Physik komplexer Systeme,\\
N\"othnitzer  Str. 38, D-01187 Dresden, Germany  } 

\maketitle

\begin{abstract}

In experiments, the dynamical behavior of systems is reflected 
in time series.
Due to the finiteness of the observational data set it is not possible
to reconstruct the invariant measure up to arbitrarily fine resolution
and arbitrarily high embedding dimension. These restrictions limit our ability to distinguish between
signals generated by different systems, such as regular, chaotic or stochastic
ones, when analyzed from a time series point of view.  We propose to classify
the signal behavior, without referring to any specific model, as stochastic or
deterministic on a certain scale of the resolution $\epsilon$, according to
the dependence of the $(\epsilon,\tau)$-entropy, $h(\epsilon, \, \tau)$, 
and of the finite size Lyapunov exponent, $\lambda(\epsilon)$, on $\epsilon$.

\end{abstract}

{PACS:05.45.Tp, 05.45.-a \hfill\break
}
%%%%%%%%%%%%%%%%%%%%%%%%%%%%%%%%%%%%%%%%%%%%%%%%%%%%%%%%%%%%%%%%%%%%%%
\section{Introduction}
\label{sec:1}
%%%%%%%%%%%%%%%%%%%%%%%%%%%%%%%%%%%%%%%%%%%%%%%%%%%%%%%%%%%%%%%%%%%%%%
It is a long debated question if and by what means we can distinguish
whether an observed irregular signal is deterministically chaotic or
stochastic
\cite{Nicolis84,Osborne89a,Sugihara90a,Casdagli91,Kaplan92a,Kubin95}. If the signal was
obtained by iterating a certain model on a computer, we can give a definite answer, because we know the law which
generated the signal. \\ In the
case of time series recorded from experimental measurements, we are in
a totally different situation.  Indeed, in most of cases, there is no
unique model of the ``system'' which produced the data. Moreover, we
will see that knowing the character of the model might not be an
adequate answer to the question of the character of the signal. For example,
data of Brownian motion can be modeled by a deterministic regular
process as well as by a deterministic chaotic or stochastic process,
as we will show in section \ref{sec:4}. \\ 
In principle, if we were able to determine the maximum Lyapunov
exponent ($\lambda$) or the Kolmogorov-Sinai entropy per unit time
($h_{KS}$) of a data sequence, we would know, with no uncertainty, whether
the sequence is generated by a deterministic law (in which case
$\lambda , h_{KS} < \infty$) or by a stochastic law (in which case
$\lambda ,h_{KS} \to \infty$). \\ In spite of their conceptual relevance, 
there are evident practical
problems with such quantities that are defined as infinite time
averages taken in the limit of arbitrary fine resolution, since, 
typically, we  have access only to a finite (and
often very limited) range of scales.  In order to cope
with these limitations, in this  paper we make use of  the ``finite
size Lyapunov exponent'' (FSLE) \cite{Aurell96}, a variant 
of the maximum Lyapunov exponent, and 
the $(\epsilon , \tau)$-entropy per unit time \cite{Kolmogorov56,Shannon49,Gaspard93c}, a
generalization of the Kolmogorov-Sinai entropy per unit
time. Basically, while for evaluating $\lambda$ and $h_{KS}$ one has
to detect the properties of a system with infinite resolution, for
determining the FSLE, $\lambda(\epsilon)$, or the $(\epsilon,
\tau)$-entropy per unit time, $h(\epsilon, \tau)$, the investigation
on the system is performed at a finite scale $\epsilon$, i.e. with a
finite resolution. $\lambda(\epsilon)$ gives us the average
exponential rate of the divergence between close (on scale $\epsilon$)
trajectories of a system, and $h(\epsilon, \tau)$ is the average rate of
information needed for prediction. If properly defined,
$h(\epsilon,\tau) \stackrel{\epsilon,\tau \to 0}{\longrightarrow}
h_{KS}$  and  $\lambda(\epsilon) \stackrel{\epsilon \to
0}{\longrightarrow} \lambda$, if $\lambda \geq 0$. Thus, if we have
the possibility of determining the behavior of $\lambda(\epsilon)$ or
$h(\epsilon,\tau)$ for arbitrarily small scales, as pointed out above,
we could answer the original question about the character
(deterministic or stochastic) of the law that generated the recorded
signal.  \\ However, the limits of infinite time
and resolution, besides being unattainable when dealing with
experimental data, may also result to be physically uninteresting.  As
a matter of fact, it is now clear that the maximum Lyapunov exponent
and the Kolmogorov-Sinai entropy are not completely satisfactory for a
proper characterization of the many faces of complexity and
predictability of nontrivial systems, such as, for instance,
intermittent systems \cite{Benzi85} or systems with many degrees of freedom
\cite{Aurell96,Grassberger89}.  For
example, in the case of the maximum Lyapunov exponent, one has to
consider infinitesimal perturbations, i.e.  infinitesimally close
trajectories resp. infinite resolution.  In
systems with many degrees of freedom (e.g. turbulence) an
infinitesimal perturbation means, from a physical point of view, that
the differences $\delta x_k =x'_k -x_k $ of the components, $x'_k$ and
$x_k$, of the initially close state vectors ${\bf x}'$ and ${\bf x}$,
have to be much smaller than the typical values $\tilde x _k $ of the
variables $x_k$. If the $\tilde x _k$-s take very different values,
then the concept of infinitesimal perturbation becomes physically
unimportant, in the case one is interested only in the evolution of
the components with the largest typical values
\cite{Lorenz69,Aurell96} (e.g., the large scales in a
turbulent motion).

Taking into account all the limitations mentioned above, in particular
the practical impossibility to reach arbitrarily fine resolution, we propose 
a different point of view on the distinction between chaos and noise. 
Neither it relies on a particular model for a given
data set nor it ignores the fact that the character of a
signal may depend on the resolution of the observation.
Indeed $h(\epsilon, \tau)$ (or equivalently $\lambda(\epsilon)$) 
usually displays different behaviors as the range of scales is varied. 
According to these different behaviors, as it will become clear 
through the paper, one can define a notion of deterministic and stochastic
behavior, respectively, on a certain range of scales.

In section \ref{sec:2} we recall
the definitions of the $(\epsilon, \tau)$-entropy and the finite size
Lyapunov exponent. In section \ref{sec:3} we discuss how one can
consistently classify the stochastic or chaotic character of a signal
by using the information theoretic concepts as the $(\epsilon,
\tau)$-entropy and the redundancy and compare our approach with
previous attempts. In section \ref{sec:4} we discuss
some examples showing that systems at the opposite in the realm of
complexity can give similar results when analyzed from a time series
point of view. Section \ref{sec:5} is devoted to a critical discussion
of some recent, intriguing and (sometimes) controversial results for
data analysis of ``microscopic'' chaos, in particular we comment
the point of view to be adopted in interpreting the result of
sec. \ref{sec:4}. In section \ref{sec:6} the reader finds
some remarks on non trivial behaviors of high dimensional systems.
Section \ref{sec:7} summarizes and concludes the paper.

%%%%%%%%%%%%%%%%%%%%%%%%%%%%%%%%%%%%%%%%%%%%%%%%%%%%%%%%%%%%%%%%%
\section{Two concepts for a resolution dependent time series analysis}
\label{sec:2}
%%%%%%%%%%%%%%%%%%%%%%%%%%%%%%%%%%%%%%%%%%%%%%%%%%%%%%%%%%%%%%%%%
\subsection{$(\epsilon, \tau)$-entropy and redundancy}
\label{sec:2.1}
%%%%%%%%%%%%%%%%%%%%%%%%%%%%%%%%%%%%%%%%%%%%%%%%%%%%%%%%%%%%%%%%%

In this section we recall the definition of the $(\epsilon, \tau)$-entropy
discussing its numerical computation, the possible technical problems
as well as its properties.

We start with a continuous (in time) variable  ${\bf x}(t) \in {\rm I}\!{\rm R}^d$,
which represents the state of a $d$-dimensional system, we discretize 
the time by introducing a time interval $\tau$ and we consider the new
variable
\begin{equation}
\label{eq:2-1}
{\bf X}^{(m)}(t)= \left( {\bf x}(t), {\bf x}(t+\tau), \dots, 
{\bf x}(t+m\tau-\tau) \right). 
\end{equation}
Of course ${\bf X}^{(m)}(t) \in {\rm I}\!{\rm R}^{md}$ and it corresponds to 
the discretized trajectory in a time interval $T=m \tau$. \\
Usually, in data analysis, the space where the state vectors of the
system live is not known. Mostly, only a scalar
variable $u(t)$ can be measured. In these cases 
one considers vectors ${\bf X}^{(m)}(t)= \left( u(t),
u(t+\tau), \dots, u(t+m\tau-\tau) \right)$, that live in ${\rm I}\!{\rm R}^m$
and allow a reconstruction of the original phase space, known as delay
embedding in the literature \cite{Takens80,Sauer91a}. It can be viewed as a
special case of (\ref{eq:2-1}). 

We introduce now a partition of the phase space ${\rm I}\!{\rm R}^d$, using
cells of length $\epsilon$ in each of the $d$ directions.  Since the
region where a bounded motion evolves contains a finite number of
cells, each ${\bf X}^{(m)}(t)$ can be coded into a word of length $m$, out of a finite alphabet:
\begin{equation}
\label{eq:2-2}
{\bf X}^{(m)}(t) \longrightarrow W^{m}(\epsilon, t) =
\left( i(\epsilon, t), i(\epsilon, t+\tau), \dots, 
i(\epsilon, t+m \tau -\tau) \right), 
\end{equation}
where $i(\epsilon, t+j \tau)$ labels the cell in ${\rm I}\!{\rm R}^d$
containing ${\bf x}(t+j \tau)$. From the time evolution of ${\bf
X}^{(m)}(t)$ one obtains, under the hypothesis of stationarity, the
probabilities $P(W^{m}(\epsilon))$ of the admissible words $\lbrace
W^{m}(\epsilon) \rbrace$. We can now introduce 
the $(\epsilon , \tau)$-entropy per unit time, $h(\epsilon , \tau)$
\cite{Shannon49}:
\begin{eqnarray}
\label{eq:2-3a}
h_m(\epsilon , \tau)&=& {1 \over \tau} \lbrack H_{m+1} (\epsilon,\tau)
-H_m (\epsilon,\tau) \rbrack \\
\label{eq:2-3b}
h(\epsilon , \tau) &=& \lim _{m \to \infty} h_m(\epsilon , \tau) = {1 \over \tau} 
\lim _{m \to \infty} {1 \over m} H_{m} (\epsilon,\tau) ,
\end{eqnarray} 
where $H_m$ is the block entropy of block length $m$:
\begin{equation}
\label{eq:2-4}
H_{m} (\epsilon,\tau) = - \sum _{ \lbrace W^{m}(\epsilon) \rbrace } 
P(W^{m}(\epsilon))  \ln P(W^{m}(\epsilon)).
\end{equation} 
For sake of simplicity, we ignored the dependence on details of the
partition. 
For a more rigorous definition one has to take into account
all partitions with elements of size smaller than
$\epsilon$, and then define $h(\epsilon,\tau)$ by the infimum over all
these partitions (see e.g. \cite{Gaspard93c}). In numerical calculations we circumvent
this difficulty by using coverings instead of partitions (see below).

A concept which is complementary to the $\epsilon$-entropy is the 
$\epsilon$-redundancy (see e.g. \cite{Fraser89}) which 
measures the amount of uncertainty on future observations which can 
be removed by the knowledge of the past, namely:
\[
r_m(\epsilon,\tau)={1 \over \tau} \left[ H_1(\epsilon,\tau)
-(H_{m+1}(\epsilon,\tau)-H_m(\epsilon,\tau,)) \right],
\]
where $H_1(\epsilon)$ estimates the uncertainty of the single outcome of the 
measurement, i.e. neglecting possible correlations in the
signal. Alternatively, we can write the redundancy in the form
\begin{equation}
r_m(\epsilon,\tau)={1 \over \tau} H_1(\epsilon,\tau)
- h_m(\epsilon,\tau) \; ,
\label{redundancy}
\end{equation}
which emphasizes the complementarity between redundancy and entropy.
If the data are totally independent, i.e. IID, one has
$H_m(\epsilon,\tau)=mH_1(\epsilon)$
and, therefore, $r_m(\epsilon,\tau)=0$. On the opposite side, in the case of a
periodic signal the redundancy is maximal $r_m(\epsilon,\tau)=H_1(\epsilon,\tau)/\tau$.

The Kolmogorov-Sinai entropy, $h_{KS}$, is obtained by
taking the limit $\epsilon,\tau \to 0$:
\begin{equation}
\label{eq:2-5}
h_{KS} = \lim _{\tau \to 0} \lim_{\epsilon \to 0} h(\epsilon, \tau). 
\end{equation} 
The KS-entropy is a dynamical invariant, i.e. it is independent of the
used state representation (\ref{eq:2-1}), while
 this is not the case for the $\epsilon$-entropy
(\ref{eq:2-3b}). To simplify the notation we drop the $\tau$ dependence
in the following, apart from cases in which the $\tau$ dependency is
explicitly considered as in section IV.

In a genuine deterministic chaotic system one has
$0< h_{KS} < \infty $ ($h_{KS}=0$ for a regular motion), while for a
random process $h_{KS}=\infty$.  The entropies $H_m(\epsilon)$
were above introduced using a partition and the usual Shannon entropy,
however it is possible to arrive at the same notion starting from other
entropy-like quantities, which are more suitable for numerical
investigations. Following Cohen and Procaccia \cite{Cohen85} one can
estimate $H_m(\epsilon)$ as follows.  Given a signal composed of $N$
successive records and the embedding variable ${\bf X}^{(m)}$, let us
introduce the quantities: 
\begin{equation} 
n_j^{(m)}={1 \over N-m-1} \sum_{i \neq
j} \Theta(\epsilon-|{\bf X}^{(m)}(i \tau)-{\bf X}^{(m)}(j \tau)|)\, ,
\end{equation}
then the block entropy $H_m(\epsilon)$ is given by 
\begin{equation}
H_m^{(1)}(\epsilon)=-\frac{1}{ (N-m)} \sum_j \ln n_j^{(m)}(\epsilon)\,.
\label{H-CP}
\end{equation}
In practice $n_j^{(m)}(\epsilon)$ is an approximation of $P(W^m(\epsilon))$.
From the numerical point of view the
even more suited quantities are the correlation 
entropies \cite{Grassberger83c,Takens98} 
\begin{equation}
H_m^{(2)}(\epsilon)=- \ln \left({1 \over N-m} \sum_j
n_j^{(m)}(\epsilon)\right) \leq H_m^{(1)}(\epsilon)\,, 
\end{equation} 
where one approximates the Shannon entropy by the 
Renyi entropy  of order $q=2$.

In the determination of $h_{KS}$ by data analysis,
one has to consider some subtle points (see \cite{Kantz97c} for a
detailed discussion). \\
Let us just make some remarks about the general problems in the
computation of the Kolmogorov-Sinai entropy from a time series of a
deterministic system. The first point is the value of the embedding
dimension $m$. Let us assume that the information dimension of the
attractor of the deterministic system is $D$.
In order to be able to observe a finite entropy, $m$ has to be larger
than $D$, since the behavior of the entropies in the limit $\epsilon
\to 0$ is  
\begin{equation}
h_m(\epsilon) = const +O( \epsilon) \ge h_{KS}\; ,
\label{deterministic_h}
\end{equation}
provided $m > D$ \cite{Ding92}.
The second relevant point is the fact that the saturation, i.e. the
regime where the entropy $h_m(\epsilon)$ does not depend on the length
scale $\epsilon$, can be observed only on length scales smaller than some
$\epsilon_u$. Thus  it
is possible to distinguish a deterministic signal from a random one
only for $\epsilon<\epsilon_u$. Due to the finiteness of the data set
there is a lower scale $\epsilon_l$ below which no information can be
extracted from the data.
Taking into account the number of points of the series, $N$, it is
possible to give the following relation 
between the embedding dimension, the $KS$-entropy, the information 
dimension and the saturation range $\epsilon_u/\epsilon_l$ \cite{Olbrich97}:
\begin{equation}
\frac{\epsilon_u}{\epsilon_l} \le
\left( N e^{-m \tau h_{KS}}\right)^{1 \over D}\; ,
\label{sat}
\end{equation}
where $\epsilon_u$ and $\epsilon_l$ are the upper and lower bounds of
the interval of scales at which the deterministic character of a
deterministic signal shows up. Note that this relation does not
determine $\epsilon_u$. For more details see \cite{Olbrich97}. \\
If $m$ is not large enough and/or $\epsilon$ is not  small enough one can
obtain misleading results, e.g. see sec. \ref{sec:5}.

The $\epsilon$-entropy $h(\epsilon,\tau)$ is well defined 
also for stochastic processes. Its dependence on $\epsilon$ can give
some insight in the underlying stochastic process
\cite{Gaspard93c}. In the case of finite $\tau$ it is possible to
define a saturation range: below some length scale
$\epsilon_u(\tau)$ we have  
\begin{equation}
h_m(\epsilon) = const -\ln \epsilon +O( \epsilon) \;
\label{stochastic_h}
\end{equation}
But, the limit $\tau \to 0$ will lead to $\epsilon_u \to
0$, thus the saturation will disappear. As it was shown
in \cite{Gaspard93c}, for some stochastic processes it is possible to
give an explicit expression of $h(\epsilon, \tau)$ in this limit. 
For instance, in the case of a stationary Gaussian process with spectrum
$S(\omega)\propto \omega^{-2}$  one has \cite{Kolmogorov56}:
\begin{equation}
\label{eq:2-6}
\lim_{\tau\to 0} h(\epsilon,\tau) \sim {1 \over \epsilon
^2} \; ,
\label{eq:kolmo56}
\end{equation}
the same scaling behavior is also expected for Brownian motion \cite{Gaspard93c}. 
It can be recovered by looking at $h(\epsilon,\tau)$ in a certain
$(\epsilon,\tau)$ region. See Ref. \cite{Gaspard93c} for a detailed derivation
of (\ref{eq:kolmo56}).  We have to stress that the behavior predicted
by eq. (\ref{eq:kolmo56}) may be difficult to be experimentally
observed due to problems related to the choice of $\tau$ (see
sect. \ref{sec:4}).

%%%%%%%%%%%%%%%%%%%%%%%%%%%%%%%%%%%%%%%%%%%%%%%%%%%%%%%%%%%%%%%%%%%
\subsection{Finite Size Lyapunov Exponent}
\label{sec:2.2}
%%%%%%%%%%%%%%%%%%%%%%%%%%%%%%%%%%%%%%%%%%%%%%%%%%%%%%%%%%%%%%%%%

The Finite Size Lyapunov
Exponent (FSLE) was originally introduced in the context
of the predictability problem in fully developed turbulence
\cite{Aurell96}. Such an indicator, as it will become clear below, is for some
aspects the dynamical systems counterpart of the $\epsilon$-entropy.

The basic idea of the FSLE is to define a growth rate for different sizes of
the distance between a reference and a perturbed trajectory.  In the
following we discuss how the FSLE can be computed, by assuming to know
the evolution equations. The generalization to data analysis is
obtained following the usual ideas of ``embedology'' \cite{Sauer91a}.
First one has to define a norm to measure the distance
$\epsilon(t)=||\delta{\bf x}(t)||$ between a reference and a perturbed
trajectory. In finite dimensional systems the maximum
Lyapunov exponent is independent of the used norm.  But when one
considers finite perturbations there could be a dependence on the
norm (as for infinite dimensional systems). Having 
defined the norm, one has to introduce a series of thresholds starting from a
very small one $\epsilon_0$, e.g.,
$\epsilon_{n}=r^{n}\epsilon_{0}$ ($n=1,\dots, P$), and to measure the
``doubling time'' ($T_{r}(\epsilon_{n})$) at different thresholds.
$T_{r}(\epsilon_{n})$ is the time a perturbation of size
$\epsilon_{n}$ takes to grow up to the next threshold,
$\epsilon_{n+1}$.  The threshold rate $r$ should not be taken too
large, in order to avoid the error to grow through different
scales. On the other hand, $r$ cannot be too close to one, because
otherwise the doubling time would be of the order of the time step in
the integration (sampling time in data analysis) affecting the
statistics.  Typically, one uses $r=2$ or $r=\sqrt 2$. For simplicity
$T_{r}$ is called ``doubling time'' even if $r \neq 2$.

The doubling times $T_{r}(\epsilon_{n})$ are obtained by following the
evolution of the distance $||\delta{\bf x}(t)||$ from its initial value
 $\epsilon_{min} \ll \epsilon_0$ up to the largest threshold 
$\epsilon_{P}$.  Knowing the evolution equations, this is obtained
by integrating the two trajectories of the system starting at an
initial distance $\epsilon_{min}$. In general, one must choose
$\epsilon_{min} \ll \epsilon_{0}$, in order to allow the direction of
the initial perturbation to align with the most unstable direction in
the phase-space. Moreover, one must pay attention to keep
$\epsilon_{P} < \epsilon_{saturation}$, so that all the thresholds can
be attained ($\epsilon_{saturation}$ is the typical distance of two
uncorrelated trajectories).

The evolution of the error from the initial value $\epsilon_{min}$ to
the largest threshold $\epsilon_{P}$ carries out a single
error-doubling experiment. At this point  the model
trajectory is rescaled at a distance $\epsilon_{min}$ with respect to
the true one, and another experiment starts out.  After ${\cal N}$
error-doubling experiments, we can estimate the expectation value of
some quantity $A$ as:
\begin{equation}
\langle A \rangle_{e} = {1 \over {\cal N}} \sum_{i=1}^{\cal N} \, A_i \, .
\label{eq:ap1}
\end{equation}
This is not the same as taking the time average because each error
doubling experiment may take a different time from the others.
For the doubling time we have
\begin{equation}
\lambda(\epsilon_n) = {1 \over \langle T_{r}(\epsilon_n) 
\rangle_{e}} \ln r \, ;
\label{eq:ap3}
\end{equation}
for details, see \cite{Aurell96}.
The method described  above assumes that the distance between the two
trajectories is continuous in time. This is not the case for maps or for
discrete sampling in time, thus the method has to be slightly
modified.  In this case $T_{r}(\epsilon_n)$ is defined as the minimum
time at which $\epsilon(T_r) \ge r \epsilon_n$, and now we have \cite{Aurell96}
\begin{equation}
\lambda(\epsilon_n) = {1 \over \langle T_{r}(\epsilon_n) \rangle_{e}}
\left\langle \ln \left( {\epsilon(T_r) \over \epsilon_n} \right) 
\right\rangle_{e} \,.
\label{eq:ap4}
\end{equation}

It is worth to note that the computation of the FSLE is not more
expensive than the computation of the Lyapunov exponent by the standard
algorithm. One has simply to integrate two copies of the system and
this can be done also for very complex simulations.  

One can expect that in systems with only one positive Lyapunov
exponent, one has $\lambda (\epsilon) \simeq h(\epsilon)$, see
Ref.\cite{Aurell96} for details. Additionally it is shown  in
Ref. \cite{Boffetta98} how it is possible to use
the FSLE for characterizing  the predictability (also
from the data analysis point of view) of systems containing a slow
component and a fast one. Let us comment on some advantages of the
FSLE with respect to the $(\epsilon , \tau)$-entropy. For the FSLE it
is not necessary to introduce an $\epsilon$-partition and, most
important, at variance with the $(\epsilon , \tau)$-entropy, the
algorithmic procedure automatically finds the ``proper time''; so that
it is not necessary to decide on the right sampling time and to test
the convergence at varying the words block size $N$. This point will
be discussed in section \ref{sec:4.2}.

%%%%%%%%%%%%%%%%%%%%%%%%%%%%%%%%%%%%%%%%%%%%%%%%%%%%%%%%%%%%%%%%
\section{Classification by $\epsilon$-dependence}
\label{sec:3}
%%%%%%%%%%%%%%%%%%%%%%%%%%%%%%%%%%%%%%%%%%%%%%%%%%%%%%%%%%%%%%%%

In the previous section we discussed the $\epsilon$-entropy and the
FSLE as tools to characterize dynamical processes. Let us re-examine
the question of distinguishing chaos and noise posed in the Introduction.
Equations  (\ref{deterministic_h}) and (\ref{stochastic_h})
allow us to make rigorous statements about the behavior of the
entropy in the limit $\epsilon \to 0$. Then the behavior of the redundancy
can be determined by using the relation to the entropy, given by
(\ref{redundancy}) if we take into account that $H_1(\epsilon) \propto
-\ln \epsilon$ for continuous valued non-periodic process. Both the
behavior of the entropy and the redundancy are summarized in the
following table \cite{Kubin95}.
\vglue5mm 
\centerline{
\begin{tabular}[h]{|c|c|}
\hline
deterministic ($m >D$) & stochastic  \\
\hline 
$r_m(\epsilon) \to \infty$ & $h_m(\epsilon) \to \infty$ \\
\hline
\begin{tabular}{c|c} 
chaotic & non-chaotic \\
\hline
$\lim_{m \to \infty} h_m(\epsilon)> 0$ &  
$\lim_{m\to \infty}h_m(\epsilon) = 0$ 
\end{tabular} & \begin{tabular}{c|c}
white noise & colored noise \\
\hline 
$r_m(\epsilon)=0$ & $r_m(\epsilon) > 0$ 
\end{tabular} \\ \hline 
\end{tabular}}
\vglue5mm
The behavior of the FSLE in the limit $\epsilon \to 0$ is similar to that of 
the $\epsilon$-entropy, $h(\epsilon)$. It is worth noting that the
FSLE defined through the doubling times (see sect. \ref{sec:2.2}) is zero 
also if $\lambda<0$.

\begin{figure}
\centerline{\psfig{file=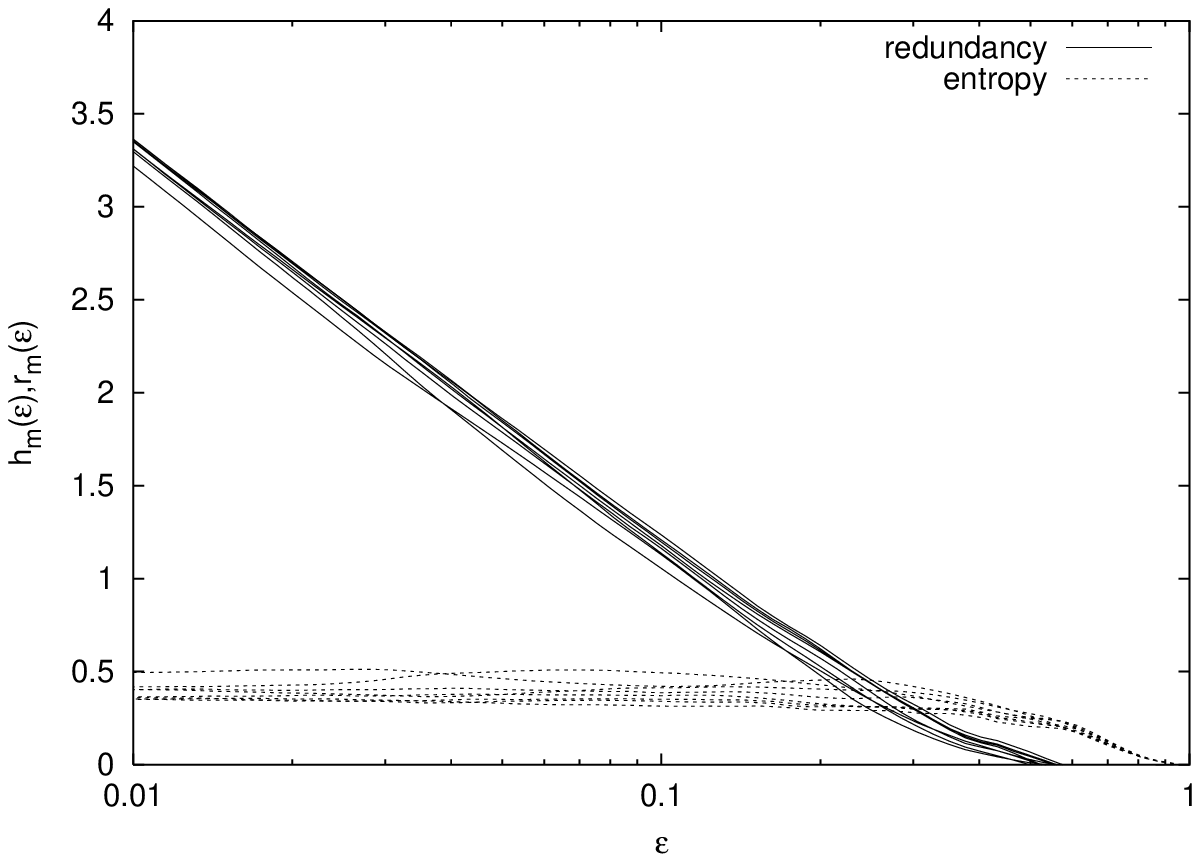,width=8cm,angle=0}
\psfig{file=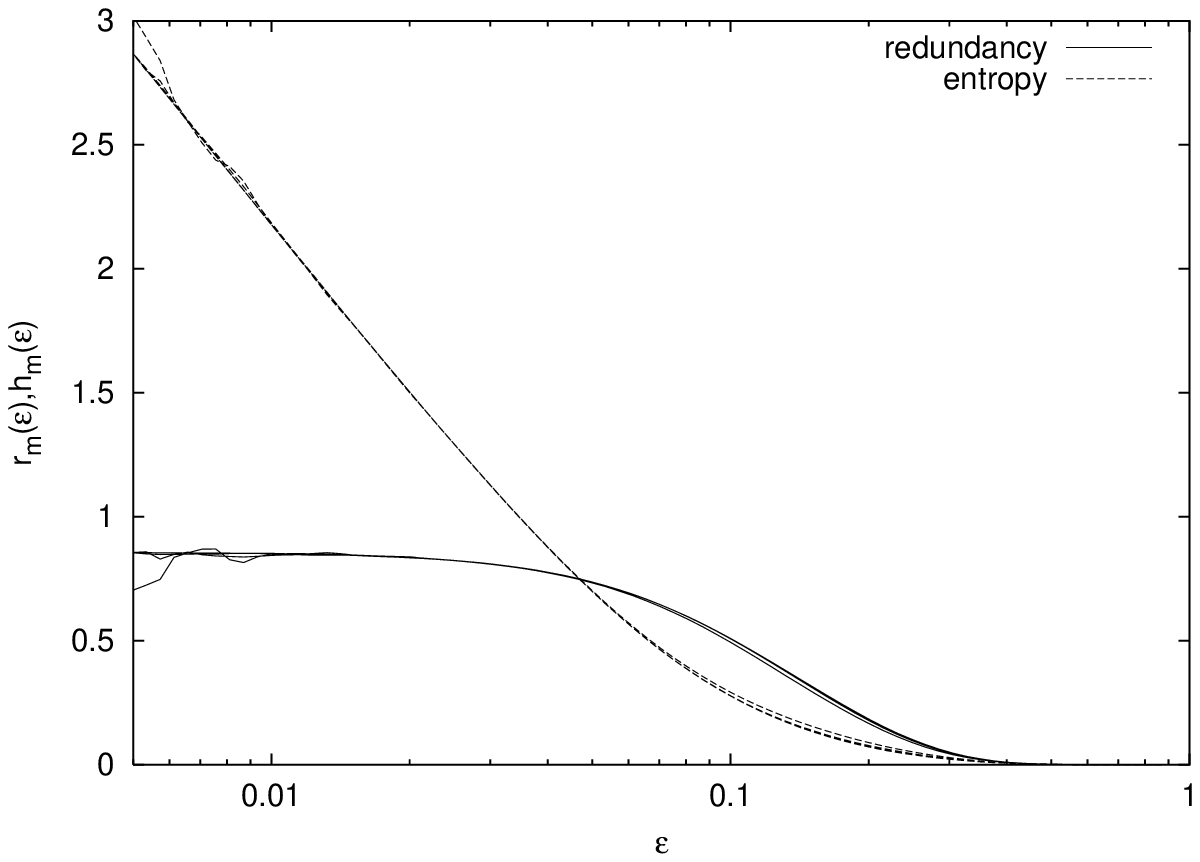,width=8cm,angle=0}}
\centerline{(a) \hglue7.5cm (b)}
\vspace{.5cm}
\caption{(a) $h_m(\epsilon)$ -dashed - and $r_m(\epsilon)$ - solid -
for the Henon map with the standard parameters ($a=1.5$ and $b=0.3$), 
with $m=2,\ldots,9$ and (b) the same for an AR(1) process, with
$m=1,\ldots,5$ and fixed $\tau$.}
\label{typical_r_h}
\end{figure}

In all practical situations we have only a finite
amount of data.  Let us assume we have embedded the time series in a
$m$-dimensional space, e.g. by time delay embedding.  Then one can
relate to this set of points an empirical measure $\mu^*$ 
\be
\mu^*({\bf X}^{(m)})=\frac{1}{N} \sum_{i=1}^N 
\delta({\bf X}^{(m)}-{\bf X}^{(m)}(i \tau)) \; .
\ee 
This empirical measure $\mu^*$ approximates the true measure only
on length scales larger than a finite length scale $\epsilon_l$. This
means that we cannot perform the limit $\epsilon \to 0$. Of course on
a finite scale $\epsilon_l$, both entropy and redundancy are always
finite, therefore we are unable to decide which will reach infinity
for $\epsilon \to 0$. But we can define stochastic and
deterministic behavior of a time series at the length
scale dependence of the entropy and the redundancy.
Fig.~\ref{typical_r_h} shows the typical behavior of the entropy
$h_m(\epsilon)$ and the redundancy $r_m(\epsilon)$ in case of a
deterministic model (2-dimensional chaotic map) and a
stochastic model (autoregressive model of first order).

For a time series  long enough, a ``typical'' system can show a
saturation range for both the entropy and the redundancy. For decreasing length
scales $\epsilon$ with $\epsilon < \epsilon_u$ one observes the following behaviors
\vglue5mm
\centerline{\begin{tabular}[h]{|c|c|}
\hline
deterministic & stochastic  \\
\hline 
$r_m(\epsilon) \propto -\ln \epsilon$  & 
$h_m(\epsilon) \propto -\ln \epsilon$ \\
\hline
$h_m(\epsilon) \approx const$ & $r_m(\epsilon) \approx const$ \\
\hline 
\end{tabular}}
\vglue5mm 
In addition, as far as stochastic behaviors are concerned, the 
$\epsilon$-entropy
can exhibit power laws  on large scales, e.g. in the case of diffusion
eq.~(\ref{eq:kolmo56}) (see section IV and Ref.~\cite{Gaspard93c} for
further details).

If on some range of length scales either the entropy $h_m(\epsilon)$
or the redundancy $r_m(\epsilon)$ is a constant, we call the signal
deterministic or stochastic on these length scales, respectively. 
Thus we have a practical tool to classify the character of a signal as
deterministic or stochastic without referring to a model, and we are no
longer obliged to answer the metaphysical question, whether the system
which produced the data was a deterministic or a stochastic one. \\
Moreover, from this point of view, we are now able to give the notion
of noisy chaos a clear meaning: chaotic scaling on large scales,
stochastic scaling on small scales. We have also the notion of chaotic
noise, namely stochastic scaling on large scales and deterministic
scaling on small scales. These notions will become clear with the
examples in the following sections.

The presented method for distinguishing between chaos and noise is a
refinement and generalization of one of the first discussed methods to
approach this problem: estimating the correlation dimension and taking a
finite value as a sign for the deterministic nature of the
signal \cite{Grassberger83d}. The main criticism to this approach is based on the
work of Osborne and Provenzale \cite{Osborne89a}, where they claimed
that stochastic systems with a power-law spectrum will produce time
series which exhibit a finite correlation dimension. 
  A detailed discussion of this problem is beyond the scope
of this paper, but a main step to clarify the problem was taken by
Theiler \cite{Theiler91}. Firstly, he noted that the discussed signals
were non-stationary and highly correlated with correlation times of
the order of the length of the time series. From a conservative point
of view one has to stop at this point with any attempt to calculate
dimensions or entropies. If one proceeds nonetheless, Theiler showed
that the result will depend on the number of data points and the
length scale. If one has a sufficient number of data points one will
encounter also for this kind of signals the embedding dimension which
leads to the typical behavior as given in (\ref{stochastic_h}) for
the entropy. Moreover, if one uses the usual time delay embedding like
(\ref{eq:2-1}) in contrast to \cite{Osborne89a} and \cite{Theiler91},
the result depends strongly on the chosen delay time.    
  
We are aware that there are a lot of other attempts to distinguish
chaos from noise discussed in the literature. They are based on the
difference in the predictability using prediction algorithms rather than
the estimating the entropy \cite{Sugihara90a,Casdagli91} or they 
relate determinism to the
smoothness of the signal \cite{Kaplan92a,Kaplan93c}. All these methods
have in common that one has to choose a certain length scale
$\epsilon$ and a particular embedding dimension $m$. 
Thus they could, in principle, shed light at the interesting crossover
scenarios we are going to describe in the next chapter.

%%%%%%%%%%%%%%%%%%%%%%%%%%%%%%%%%%%%%%%%%%%%%%%%%%%%%%%%%%%%%%%%
\section{Difficulties in the distinction between chaos and noise: examples}
\label{sec:4}
%%%%%%%%%%%%%%%%%%%%%%%%%%%%%%%%%%%%%%%%%%%%%%%%%%%%%%%%%%%%%%%%

In this section we analyze in a detailed way some examples
which illustrate how subtle the transition from the large scales behavior
to the small scales one can be; and thus the  difficulties arising 
in distinguishing, only from data analysis, a genuine deterministic 
chaotic system from one with intrinsic randomness. 
%%%%%%%%%%%%%%%%%%%%%%%%%%%%%%%%%%%%%%%%%%%%%
\subsection{The diffusive regime}
\label{sec:4.1}
%%%%%%%%%%%%%%%%%%%%%%%%%%%%%%%%%%%%%%%%%%%%%

We first discuss the problems, due to the finite resolution, which 
one can have in analyzing experimental data.
We consider  the following map which generates a diffusive behavior on the
large scales \cite{Schell82}:
\begin{equation}
\label{eq:3-1}
x_{t+1} = \lbrack x_{t} \rbrack + F\left(x_{t} - \lbrack x_{t} \rbrack
 \right) ,
\end{equation} 
where $\lbrack x_{t} \rbrack$ indicates the integer part of $x_{t}$
and $F(y)$ is given by:
\begin{equation}
F(y)=\left\{\begin{array}{ll}
(2+\Delta) y & \:\: \mbox{if}\:\: y\, \in [0,1/2[ \nonumber \\
(2+\Delta) y-(1+\Delta) &\:\: \mbox{if}\:\: y\, \in ]1/2,1]\,.
\end{array}\right.
\label{eq:3-1.1}
\end{equation}
The maximum Lyapunov exponent $\lambda$ can be obtained
immediately: $\lambda = \ln |F'|$, with $F'=dF/dy=\!2\!+\!\Delta$.  Therefore
one expects the following scenario for $h(\epsilon)$ (and for $\lambda
(\epsilon))$:
\begin{equation}
\label{eq:3-2}
h(\epsilon) \approx \lambda \qquad {\rm for} \quad \epsilon < 1 ,
\end{equation} 
\begin{equation}
\label{eq:3-3}
h(\epsilon) \propto {D\over \epsilon ^2} \qquad {\rm for} 
\quad \epsilon > 1 , 
\end{equation} 
where $D$ is the diffusion coefficient, i.e. 
\begin{equation}
\label{eq:3-4}
\langle \left( x_t -x_0 \right)^2 \rangle \approx 2 \,\, D \,\, t
\qquad {\rm for} \quad {\rm large} \quad t .
\end{equation}
Fig.s~\ref{fslediff} and \ref{entrodiff} show $\lambda(\epsilon)$ and
$h(\epsilon)$ respectively. Let us briefly comment on a technical
aspect. The numerical computation of $\lambda(\epsilon)$ does not
present any particular difficulties; on the other hand the results for
$h(\epsilon)$ depend on the used sampling time $\tau$. This can be
appreciated by looking at Fig.~25b of the Gaspard and Wang review
\cite{Gaspard93c}, where the power law behavior (\ref{eq:3-3}) in the
diffusive region is obtained only if one considers the envelope of
$h_m(\epsilon,\tau)$ evaluated for different values of $\tau$; while
looking at a single $\tau$ one has a rather not conclusive result.
This is due to the fact that, at variance with the FSLE, when
computing $h(\epsilon, \tau)$ one has to consider very large $m$, in
order to obtain a good convergence for $H_{m} (\epsilon) -H_{m-1} (\epsilon)$.

Because of the diffusive behavior, a simple dimensional argument 
shows that, by sampling the system every elementary time step, 
a good convergence holds for $m \geq \epsilon ^2 / D$. 
Thus, for $\epsilon = 10$ and typical values of the diffusion 
coefficient $D \simeq  10^{-1}$, one has to consider enormous block 
size. A possible way out of this computational difficulty may be 
the following. We call $l\, (=1)$ the length of the interval $ [0,1]$ 
where $F(y)$ is defined; if we adopt a coarse-grained description 
on a scale $\epsilon=l$, i.e. we follow the evolution of the integer 
part of $x_t$, the dynamical system (\ref{eq:3-1}) is well described 
by means of a random walk \cite{Schell82}, with given probability $p_0$ that in a 
time step $s \,(=1)$ the integer part of $x$ does not change: 
$\lbrack x_{t+s} \rbrack = \lbrack x_{t} \rbrack$, and 
probabilities $p_{\pm}$ that $\lbrack x_{t+s} \rbrack = 
\lbrack x_{t} \rbrack \pm 1$. A diffusive behavior means that 
the probability for a changing of $\lbrack x_{t} \rbrack$ by 
$\pm k$ in $n$ elementary time steps is given by \cite{Feller} 
\begin{equation}
\label{eq:3-6}
P(k,n)\simeq {e^{-
                  \left( k^2 / 2\alpha n \right)  
                  \left( l^2 / s \right) 
                 }
                 \over 
                 \sqrt{2\pi \alpha n} 
                 } 
              \sqrt{ {l^2 \over s} },
\end{equation} 
with $\alpha$ a function of $p_0$. Now we increase the graining 
and identify all the $x$ values in a cell of size 
$\epsilon = L$, with $L$ an integer multiple of $l$. 
If we observe the coarse-grained state of 
the system only every $\tau (>s)$ time steps and we define the 
variables ${\tilde k}$ and ${\tilde n}$, such that 
$k l \simeq {\tilde k} L$ and $n s \simeq {\tilde n} \tau $, 
we can write 
\begin{equation}
\label{eq:3-7}
P({\tilde k}, {\tilde n}) \simeq 
             {e^{- 
             \left( {\tilde k}^2 / 2\alpha {\tilde n} \right)
             \left( L^2 / \tau  \right)
                }              
               \over 
            \sqrt{2\pi \alpha {\tilde n} } 
                } 
               \sqrt{ {L^2 \over \tau } },
\end{equation} 
for the probability of finding the system ${\tilde k}$ $L$-cells
apart after ${\tilde n}$ $\tau $-intervals. Thus, by choosing 
$L^2/\tau  = l^2/s$, we expect that the sequences generated by 
checking the system either on a scale $L$ every $\tau $ steps or 
on a scale $l$ every elementary time step $s$, have the same statistics, 
in particular the same entropy, as a signature of a diffusive behavior. 
Note that if $\lim_{m \to \infty} H_m(\epsilon, \tau =\epsilon ^2)/m$ 
is constant at varying $\epsilon$, as we found numerically, 
then the $\epsilon$-entropy per unit 
time $\lim_{m \to \infty} H_m(\epsilon, \tau )/m\tau $ goes like 
$1/\epsilon^2$. 

Since the equality $L^2/\tau = l^2/s$ is assured by the choice
$\epsilon=L=\gamma l$ and $\tau=\gamma^2 s$, one can expect that,
for a diffusion process, the following scaling relation hold: 
$\lim_{m \to \infty} H_m(\epsilon,\tau)/m  =
\lim_{m \to \infty} H_m(\gamma\epsilon,\gamma^2 \tau)/m $,
with $\gamma$ an arbitrary scaling parameter. 
This scaling relation
allows us to see why the power law behavior (\ref{eq:kolmo56}) is
expected to be valid generally for the Brownian motion.  Indeed, if we
choose $\gamma=1/\epsilon$ we have
$H_m(\epsilon,\tau)=H_m(1,\tau/\epsilon^2)$ and, finally, taking the limit
$\tau \to 0$ the $\epsilon$-entropy is given by:
\begin{equation}
h(\epsilon)=\lim_{\tau \to 0} {\left[H_{m+1}(1,\tau/\epsilon^2)-H_m(1,\tau/\epsilon^2)\right] \over \tau} \simeq
\lim_{\tau \to 0} {const\, \tau /\epsilon^2 + O(\tau^2) \over \tau}\propto {1\over
\epsilon^2}
\label{eq:added}
\end{equation}
which is (\ref{eq:kolmo56}).
Note that the first equality in (\ref{eq:added}) has been obtained by
a Taylor expansion around $\tau=0$, and by noting that $h(1,0)=0$ otherwise 
the entropy for unit time will be infinite at finite $\epsilon$ which is impossible. 
%%%%%%%%%%%%%%%%%%%%%%%%%%%%%%%%%%%%%%%%%%%%%%%%%%%%%%%%%%%%%%%%%%%%%%%
\subsection{Finite resolution effects} 
\begin{figure}[bh]
\centerline{\psfig{file=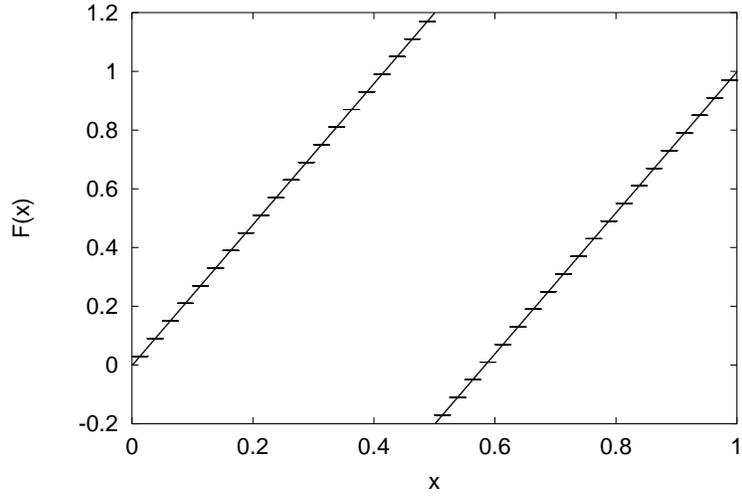,width=10cm,angle=0}}
\vspace{.5cm}
\caption{The map $F(x)$ (\ref{eq:3-1.1}) for $\Delta=.4$ is shown
with superimposed the approximating (regular) 
map $G(x)$ (\ref{eq:3-5}) obtained by using  $40$ intervals of slope $0$. }
\label{map}
\end{figure}

\begin{figure}[bh]
\centerline{\psfig{file=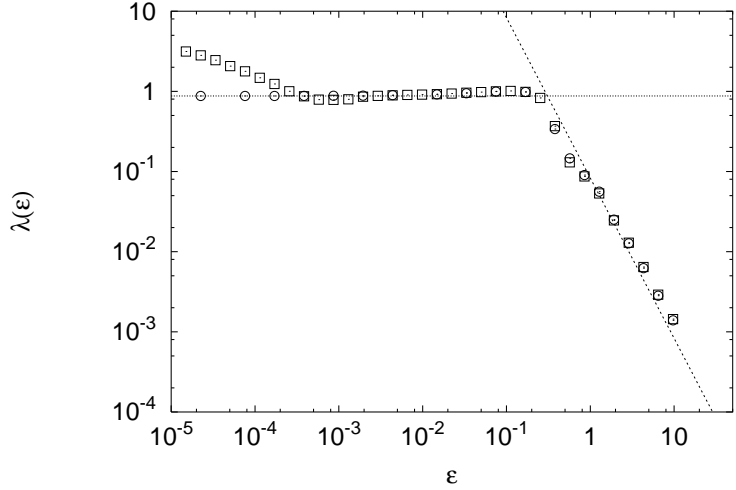,width=10cm,angle=0}}
\vspace{.5cm}
\caption{$\lambda(\epsilon)$ versus $\epsilon$ obtained with the map
$F(y)$ (\ref{eq:3-1.1}) with $\Delta=.4$ ($\circ$) and with the noisy
(regular) map ($\Box$) (\ref{eq:3-5}) with $10000$ intervals of slope $.9$ with 
$\sigma=10^{-4}$. The straight lines indicates the Lyapunov exponent
$\lambda=\ln 2.4$ and the diffusive behavior $\lambda(\epsilon) \sim \epsilon^{-2}$.}
\label{fslediff}
\end{figure}

\begin{figure}[bh]
\centerline{\psfig{file=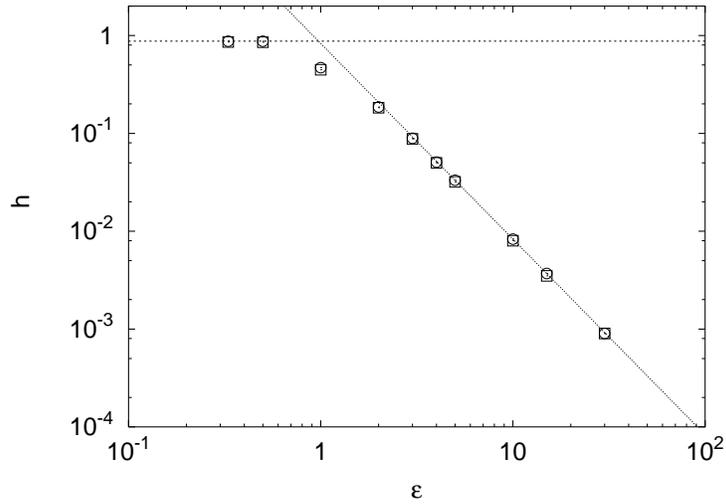,width=10cm,angle=0}}
\vspace{.5cm}
\caption{$(\epsilon,\tau)$-entropy for the noisy ($\Box$) and the 
chaotic maps ($\circ$) with the same parameters as Fig.~\ref{fslediff}, 
the encoding method is explained in the text.
The straight lines indicates the $KS$-entropy
$h_{KS}=\lambda=\ln 2.4$ and the diffusive behavior $h(\epsilon) \sim
\epsilon^{-2}$.
The region $\epsilon <\sigma$ has not be explored for computational costs.
}
\label{entrodiff}
\end{figure}

%%%%%%%%%%%%%%%%%%%%%%%%%%%%%%%%%%%%%%%%%%%%%%%%%%%%%%%%%%%%%%%%%%%%%%%
We consider now a stochastic system, namely a map with dynamical noise  
\begin{equation}
\label{eq:3-5}
x_{t+1} = \lbrack x_{t} \rbrack + G \left(x_{t} - \lbrack x_{t} \rbrack
 \right) + \sigma \eta _{t},
\end{equation} 
where $G(y)$ is shown in Fig.~\ref{map} and $\eta _{t}$ is a noise
with uniform distribution in the interval $\lbrack -1, 1 \rbrack$, and
no correlation in time. 
As it can be seen from
Fig.~\ref{map}, the new map $G(y)$ is a piecewise linear map which
approximates the map $F(y)$.  When $dG/dy < 1$, as is the case we
consider, the map (\ref{eq:3-5}), in the absence of noise, gives a
non-chaotic time evolution.

Now one can compare the chaotic case, i.e. eq. (\ref{eq:3-1})
with the approximated map (\ref{eq:3-5}) with noise.
For example let us start with the computation of the finite size Lyapunov 
exponent for the two cases. Of course from a data analysis point of view 
we have to compute the FSLE by reconstructing the dynamics by embedding.
However, for this example we are interested only in discussing the 
resolution effects. Therefore we compute the FSLE directly by integrating 
the evolution equations for two (initially) very close trajectories,
in the case of noisy maps using two different realizations of the noise.  
In Fig.~\ref{fslediff} we show $\lambda(\epsilon)$ versus $\epsilon$
both for the chaotic (\ref{eq:3-1}) and the noisy (\ref{eq:3-5}) map.
As one can see the two curves are practically indistinguishable in the region
$\epsilon >\sigma$. The differences appear only at
 very small scales $\epsilon < \sigma$ where one has 
a $\lambda(\epsilon)$ which grow with $\epsilon$ for the noisy case,
remaining at the same value for the chaotic deterministic chase.

Both the FSLE and the $(\epsilon,\tau)$-entropy analysis show that we can distinguish
three different regimes observing the dynamics of (\ref{eq:3-5}) on
different length scales. On the large length scales $\epsilon > 1$ we
observe diffusive behavior in both models. On length scales $\sigma <
\epsilon < 1$ both models show chaotic deterministic behavior,
because the entropy and the FSLE is independent of $\epsilon$ and
larger than zero. Finally on the smallest length scales $\epsilon <
\sigma $ we see stochastic behavior for the system  (\ref{eq:3-5})
while the system (\ref{eq:3-1}) still shows chaotic behavior.

%%%%%%%%%%%%%%%%%%%%%%%%%%%%%%%%%%%%%%%%%%%%%%%%%%%%%%%%%%%%%%%%%%
\subsection{Effects of finite block length}
\label{sec:4.2}
%%%%%%%%%%%%%%%%%%%%%%%%%%%%%%%%%%%%%%%%%%%%%%%%%%%%%%%%%%%%%%%%%%
In the previous section we discussed the difficulties 
arising in classifying a signal as chaotic or stochastic because of
the impossibility to reach an arbitrary fine resolution.  Here
we investigate the reasons, which make it difficult to distinguish a
stochastic behavior from a deterministic non-chaotic one. In
particular, we show that a non-chaotic deterministic system may
produce a signal practically indistinguishable from a stochastic
one, provided its phase space dimension is large enough.

The simplest way to generate a non-chaotic (regular) signal having
statistical properties similar to a stochastic one is by considering 
the Fourier expansion of a random signal \cite{Osborne89a}.
One can consider the following signal:
\begin{equation}
x(t)= \sum_{i=1}^M X_{0i}\sin \left( \Omega_i t+\phi_i \right) 
\label{eq:mazur}
\end{equation}
where  the frequencies are  equispaced discrete
frequencies, i.e. $\Omega_i=\Omega_0+i \Delta \Omega$, the phases 
$\phi_i$ are random variables uniformly distributed in 
$[0,2\pi]$ and the coefficient $X_{0i}$ are chosen in such a way
to have a definite power spectrum, e.g. a power law spectrum,
which is a common characteristic of many natural signals.
Of course (\ref{eq:mazur}) can be considered as the Fourier expansion of a
stochastic signal only if one consider a set of $2M$ points
such that $M \Delta \Omega=\pi / \Delta t$, where $\Delta t$ is the sampling time
 \cite{Osborne89a}.
Time series as (\ref{eq:mazur}) have been used to claim that suitable
stochastic signals may display a finite correlation dimension 
\cite{Osborne89a,Provenzale92}, see the discussion in sect. \ref{sec:3}. 

Here we adopt a slightly different point of view.
The signal (\ref{eq:mazur}) can also be considered as the displacement
of a harmonic oscillator linearly coupled to other harmonic
oscillators. 
Indeed,
it is well known since long times that a large ensemble of harmonic
oscillators can originate stochastic-like behaviors
\cite{Mazur60}.  In particular, we refer to~\cite{Mazur60}, where
 it was proved that an impurity of mass $\mu$ linearly coupled 
to a one-dimensional equal mass, $\mu_0$,
chain of $M$ oscillators coupled by a nearest-neighbor harmonic
interaction, in the limit of $\mu \gg \mu_0$ and of infinite
oscillators ($M \to \infty$), undergoes a Brownian motion.
Practically our observable is given by the sum of
harmonic oscillations as in eq. (\ref{eq:mazur}),
where the frequencies $\Omega_i$  have been derived in the limit
$\mu_0/\mu \ll 1$ by Cukier and Mazur \cite{Mazur60}.
The phases $\phi_i$ are chosen as uniformly distributed random
variables in $[0,2\pi]$ and the amplitudes  $X_{0i}$ are chosen 
as follows:
\begin{equation}
X_{0i}=C \Omega_i^{-1}
\end{equation}
where the $C$ is an arbitrary constant and the $\Omega$ dependence
is just to obtain a diffusive-like behavior. Note that for a signal of
length $2M$ the random phases and the $X_{0i}$'s represent a initial
condition of the $M$ oscillators, because their phase space is
$2M$-dimensional. 

\begin{figure}
\centerline{\psfig{file=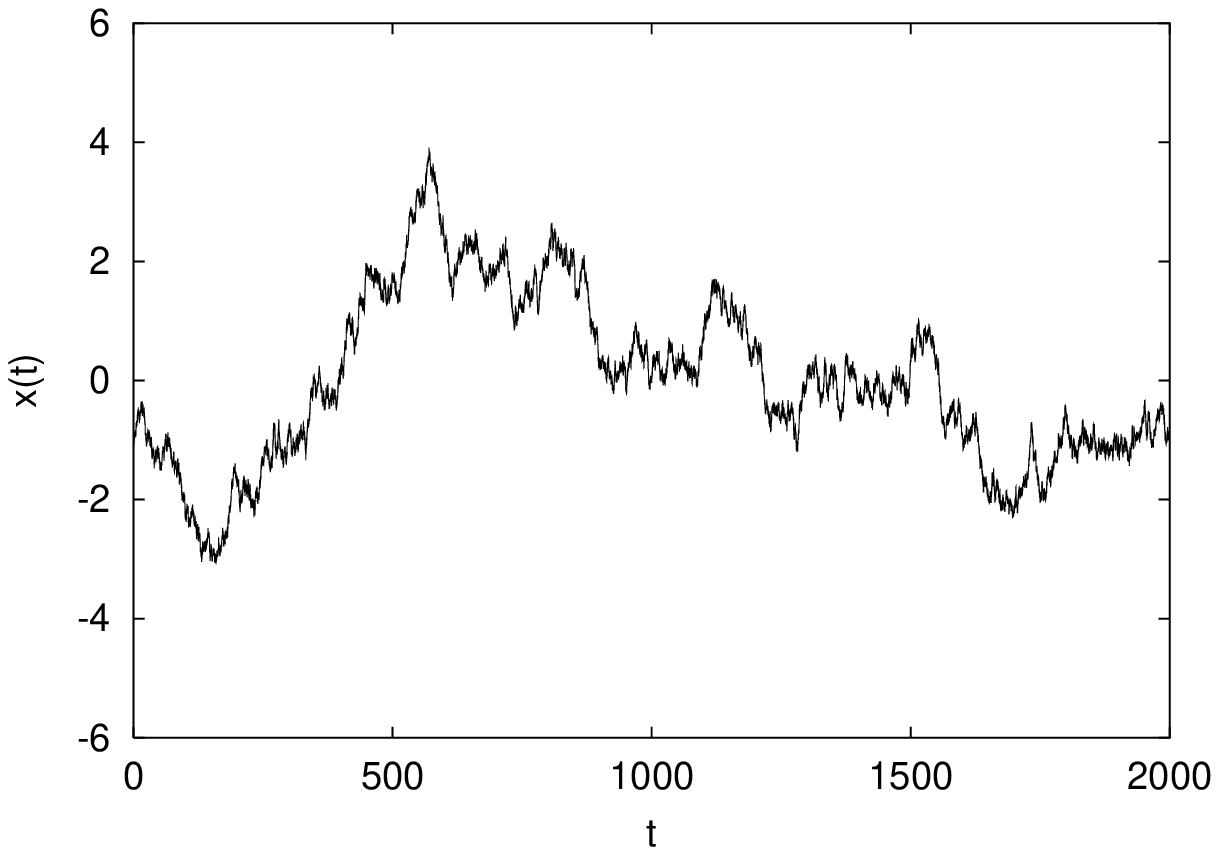,width=8cm,angle=0}
\hfill \psfig{file=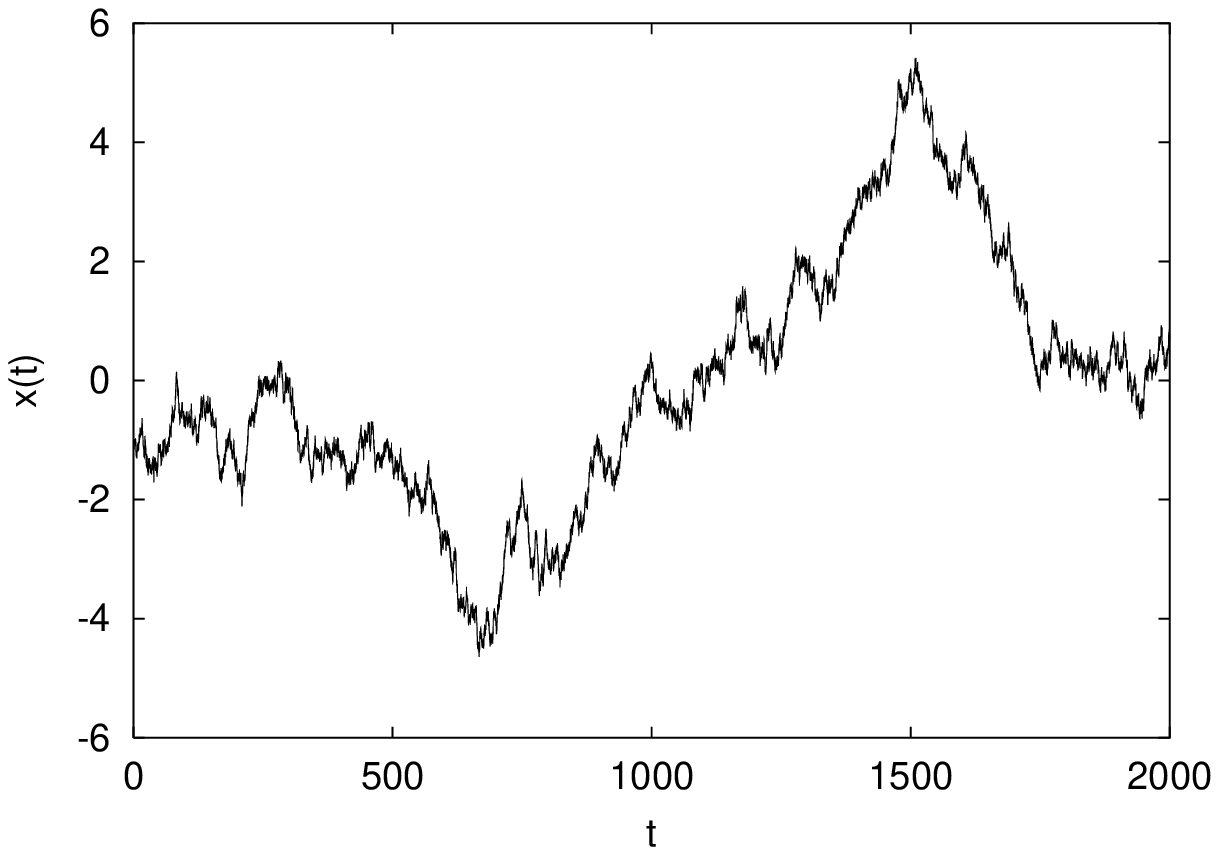,width=8cm,angle=0}}
\centerline{(a) \hglue7.5cm (b)}
\vspace{.5cm}
\caption{(a) Time record obtained from eq. (\ref{eq:mazur}) with the
frequencies chosen as discussed in the text with $M=10^4$, $C=.005$ and
$\mu_0 / \mu=10^{-6}$, the numerically computed diffusion constant is
$D \approx 0.007$. The length of the data set is $10^5$, and the data
are sampled with $\Delta t=0.02$.
(b) Time record obtained from an artificial Brownian motion (\ref{eq:brow}) 
with  the same value of the diffusion constant as in (a).}
\label{fig:signals}
\end{figure}

In Fig~\ref{fig:signals}a we show an output of the signal
(\ref{eq:mazur}) and, for comparison, in Fig~\ref{fig:signals}b we also
show an artificial continuous time Brownian motion obtained integrating the 
following equation
\begin{equation}
{{\rm d}x(t)\over {\rm d}t}=\xi(t) 
\label{eq:brow}
\end{equation}
where $\xi(t)$ is a Gaussian white noise, produced by a random number
generator (the variance of the process is chosen as to mimic that
obtained by (\ref{eq:mazur})) . Because the random number generator
uses a high entropic one-dimensional deterministic map, this is an
example for a high entropic low dimensional system, which produces
stochastic behavior. As it is possible to see the two signals appears
to be very similar already at a first sight.

\begin{figure}[h]
\centerline{\psfig{file=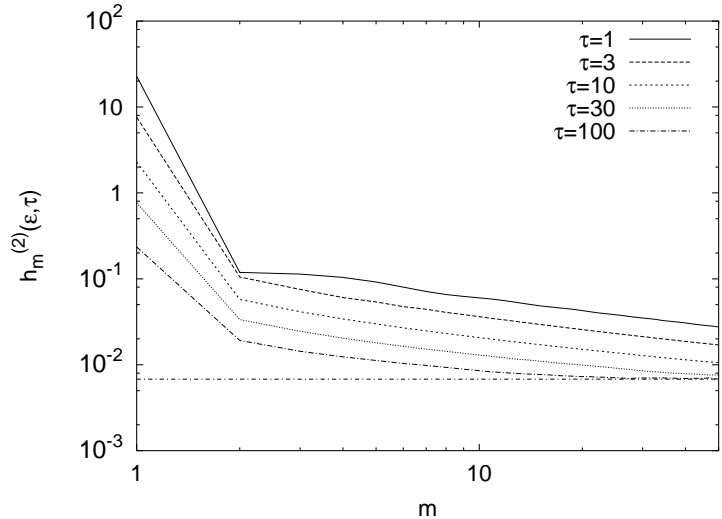,width=10cm,angle=0}}
\vspace{.5cm}
\caption{Dependence from the embedding dimension of the
$\epsilon$-entropy calculated with the Grassberger-Procaccia algorithm
using at $\epsilon=1.85$ from the time series shown in
Fig.~\ref{fig:signals}a.
The horizontal line only indicates a possible numerically evaluated value for the
saturation of the entropy.}

\label{fig:h2_of_m}
\end{figure}

Now it is important to stress that if $M<\infty$ the signals obtained 
according to (\ref{eq:mazur}) cannot develop a true Brownian motion
especially if one is interested in long time series. 
Indeed for a long enough record one should be able
to recognize the regularities in the trajectory of $x(t)$. 
However, even if the time record is long enough, in order to give a definite  
answer about the value  of the entropy one also needs very large 
embedding dimensions. The basic fact is that deterministic behavior 
can be observed only,  if the embedding dimension $m$ is larger 
than the dimension of the manifold where the
motion takes place, which is $M$ for $M$ harmonic oscillators.
This
means that although the entropy $h_{KS}$ is zero, the conditional entropies, 
$h_m(\epsilon, \tau) = (H_{m+1} (\epsilon) -H_m(\epsilon))/\tau$, 
for finite $m$ are nonzero and maybe even slowly decreasing for $m>M$.
Moreover, one can encounter some quasi-convergences with respect to $m$ for $m<M$, if 
$\tau$ is large enough, i.e. the entropy can seem to be independent of $m$,
e.g. see Fig.~\ref{fig:h2_of_m}. 

\begin{figure}[h]
\centerline{\psfig{file=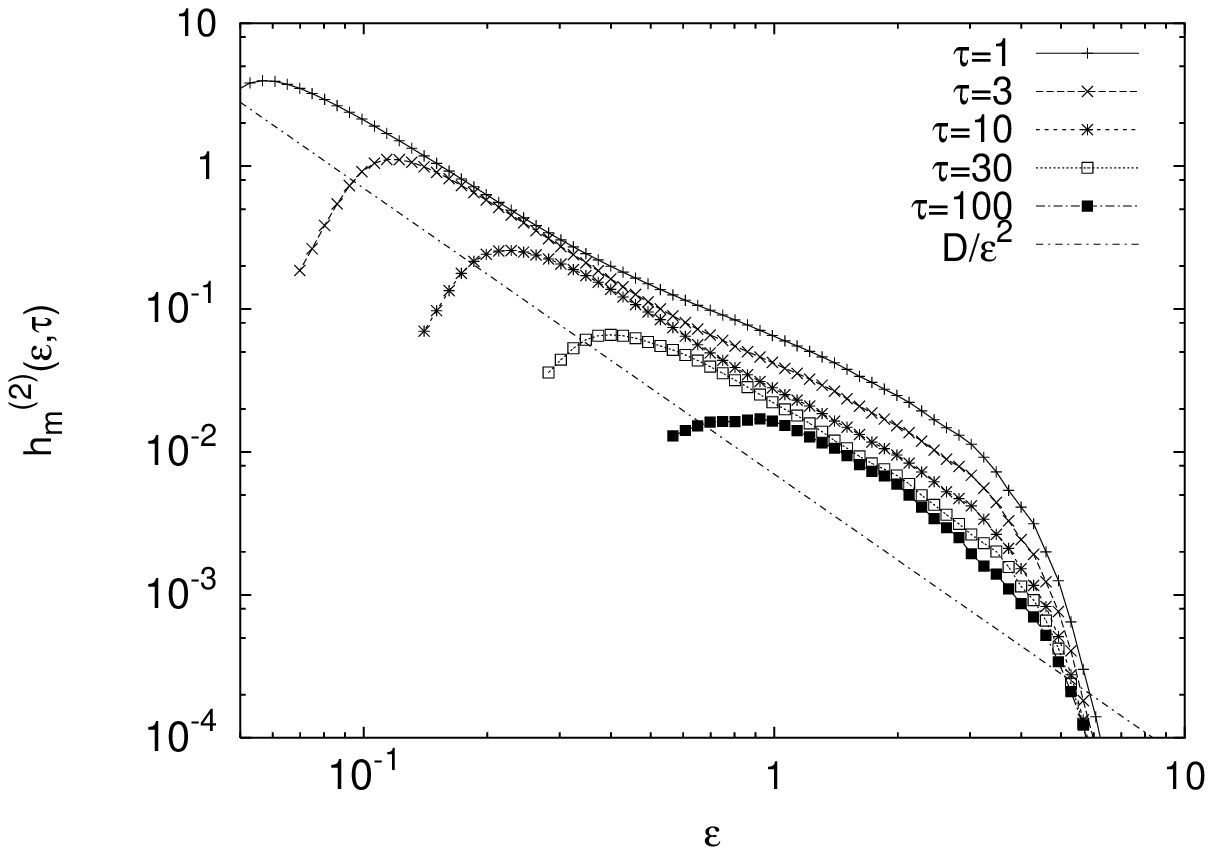,width=8cm,angle=0}
\hfill \psfig{file=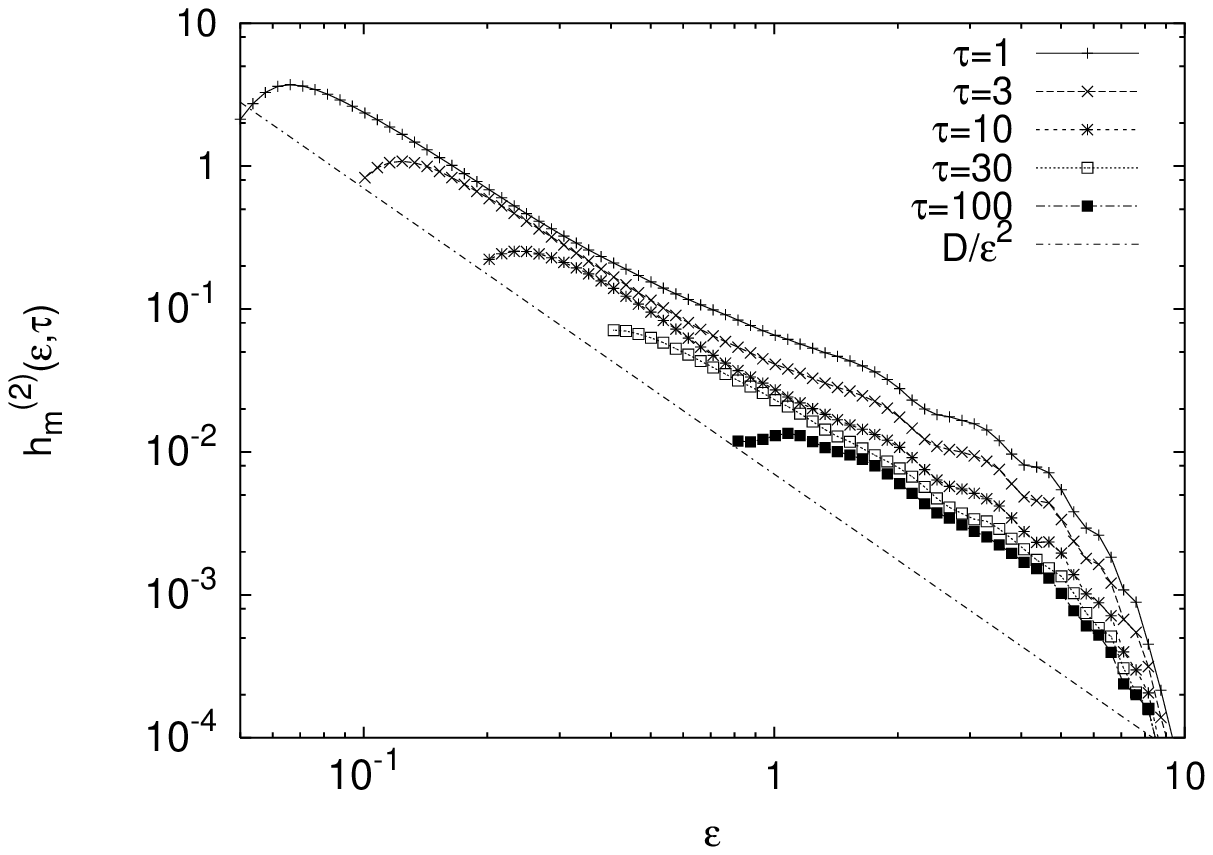,width=8cm,angle=0}}
\centerline{(a) \hglue7.5cm (b)}
\vspace{.5cm}
\caption{$\epsilon$-entropy calculated with the Grassberger-Procaccia
algorithm using using $10^5$ points from the time series shown in
Fig.~\ref{fig:signals}. We show the results for embedding dimension $m=50$.
The two straight-lines show the $D/\epsilon^2$ behavior.}
\label{fig:h2_m50}
\end{figure}

\begin{figure}[h]
\centerline{\psfig{file=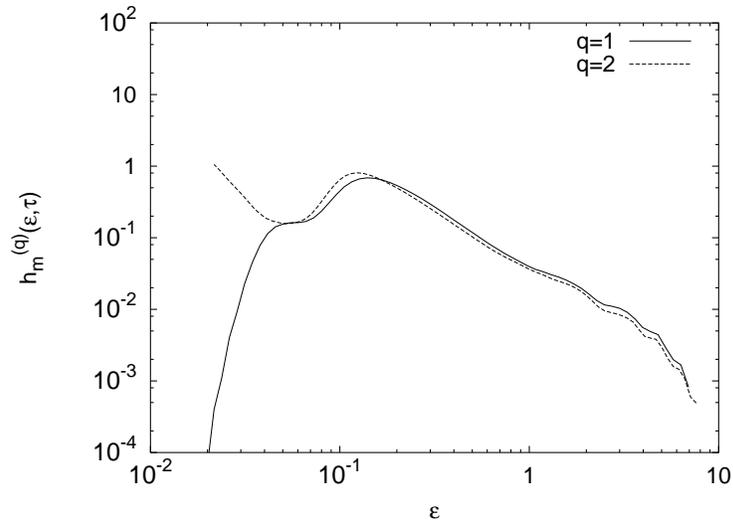,width=10cm,angle=0}}
\vspace{.5cm}
\caption{$\epsilon$-entropy estimated by the Cohen-Procaccia algorithm
($q=1$) compared to the entropies calculated by the
Grassberger-Procaccia algorithm ($q=2$) for $m=20$ and $\tau=0.2$.
Let us stress that the behavior below $\epsilon=0.2$ is essentially a finite
sample effect.}
\label{fig:compare}
\end{figure}

\begin{figure}
\centerline{\psfig{file=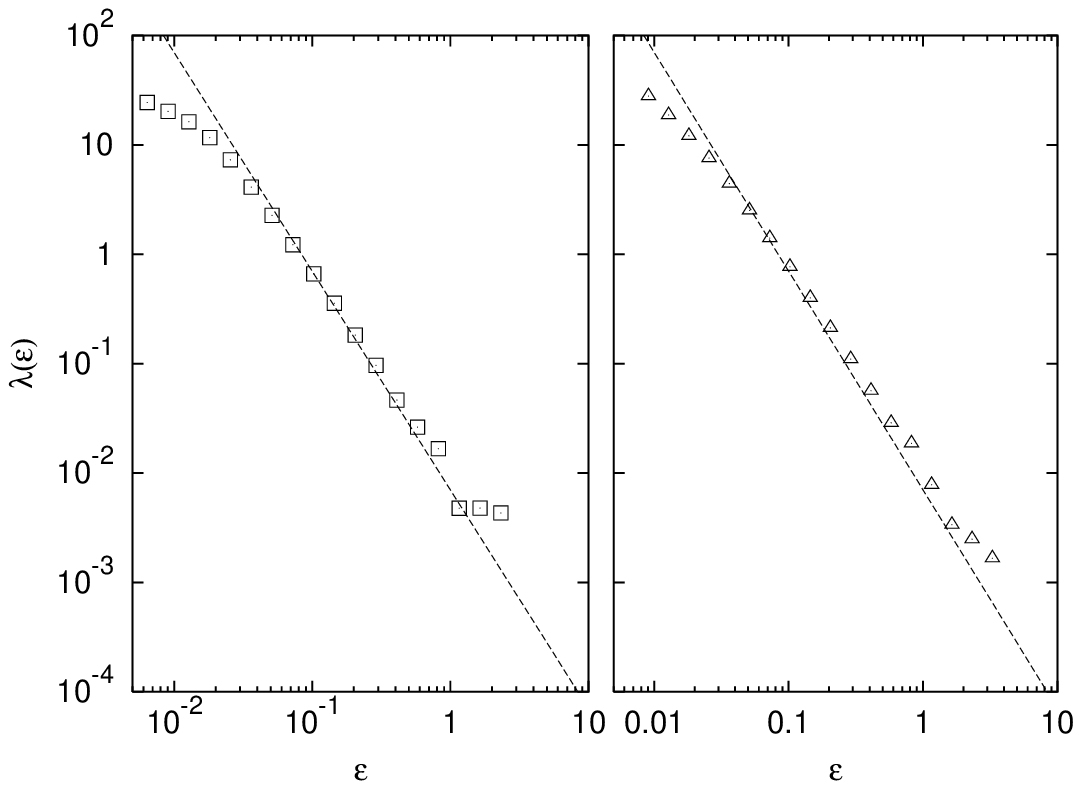,width=12cm,angle=0}}
\vspace{.5cm}
\caption{Finite size Lyapunov-Exponent from a artificial Brownian 
motion (right) and for the time series (\ref{eq:mazur}) (left). 
Embedding dimension $m=20$. Theiler window $1000$, i.e. we consider 
only neighbors which are distant in time more than $1000$ sampling 
times. The straight lines show the behavior $D/\epsilon^2$,
where $D=0.007$ is the diffusion constant. The analysis has been performed on
$10^5$ points. Different computation by changing $\tau$ and $m$ have lead to
the same result.}
\label{fig:br_mz_fsle}
\end{figure}

In Fig.~\ref{fig:h2_m50}a-b we show the
$\epsilon$-entropy, calculated by the Grassberger-Procaccia
method \cite{Grassberger83c}. The deterministic signal (\ref{eq:mazur}) and the stochastic
one (\ref{eq:brow}) produce indeed very similar results. 
Note that we calculated
$h^{(2)}_m(\epsilon,\tau)$ instead of $h^{(1)}_m(\epsilon,\tau)$
because it is used more often due to better statistics in most
cases. However, in Fig.~\ref{fig:compare} one can see, that on the relevant
length scales both entropies are almost equal.\\
As in
Refs.\cite{Gaspard93c,Gaspard98,Cohen_comment} we have considered different
time delays, $\tau$, in computing the $\epsilon$-entropy because of
the problems discussed in the previous section. The power law
behavior $\epsilon^{-2}$ for the $\epsilon$-entropy is finally
obtained only as the envelope of different computations with different
delay times. The results for the FSLE calculated from the time series
are shown in Fig.~\ref{fig:br_mz_fsle}. Both the
$\epsilon$-entropy and the FSLE display the $1/\epsilon^2$ behavior,
which denotes that the signals can be classified as Brownian motion
\cite{Gaspard93c}. 
It is worth noting that the FSLE computed from the time record 
is not too sensitive on the choice of the delay time $\tau$ and the
embedding dimension $m$.

From this simple example is it easy to understand that the impossibility of
reaching  high enough embedding dimensions severely limits our ability to make 
definite statements about the "true" character of the system which generated a
given time series as well as the already analyzed problem of the lack of resolution.

%%%%%%%%%%%%%%%%%%%%%%%%%%%%%%%%%%%%%%%%%%%%%%%%%%%%%%%%%%%%%%%%%%
\section{Some remarks on a recent debate about ``microscopic'' chaos}
\label{sec:5}
%%%%%%%%%%%%%%%%%%%%%%%%%%%%%%%%%%%%%%%%%%%%%%%%%%%%%%%%%%%%%%%%%%
The issue of the detection of ``microscopic'' chaos by data analysis
has recently received some attention
\cite{Cohen_comment,Grassberger_comment} after a work of Gaspard et
al. \cite{Gaspard98}. Gaspard et al., from an entropic analysis of an
ingenious experiment on the position of a Brownian particle in a
liquid, claim to give an empirical evidence for microscopic chaos,
i.e.: they claim to give evidence that the diffusive behavior observed
for a Brownian particle is the consequence of chaos on the molecular
scale.  Their work can be briefly summarized as follows: from a long
($\approx 1.5 \cdot 10^5$ data) record on the position of a Brownian
particle they compute the $\epsilon$-entropy with the Cohen-Procaccia
\cite{Cohen85} method,
described in section \ref{sec:2}, from which they obtain: \be h(\epsilon) \sim
{D \over \epsilon^2}
\label{eq:gasp}
\ee 
where $D$ is the diffusion coefficient. Then they {\em assume}
that the system is deterministic and therefore, because of the
inequality $h(\epsilon >0) \leq h_{KS}$, they conclude that the system
is chaotic. However, from the results presented
in the previous sections we can understand that their result does not
give a direct evidence that the system is deterministic and chaotic.
Indeed, the power law (\ref{eq:gasp}) can be produced in different ways:
\begin{enumerate}
\item a genuine chaotic system with diffusive behavior 
      (as the map (\ref{eq:3-1.1}) of sect. \ref{sec:4.1});
\item a non chaotic system with some noise, as the map (\ref{eq:3-5});
\item a genuine Brownian system, as  (\ref{eq:brow});
\item a deterministic linear non chaotic system with many degrees of
      freedom (as the example \ref{eq:mazur});
\item ``complicated'' non chaotic system as the Ehrenfest wind-tree
model where a particle diffuses in a plane due to collisions with
randomly placed, fixed oriented square scatters (as discussed by Cohen
et al. \cite{Cohen_comment} in their comment to Gaspard et
al. \cite{Gaspard98}).
\end{enumerate}
It seems to us that the very weak points of Gaspard et al. are:
\begin{description}
\item[a)] the explicit assumption that the system is deterministic
\item[b)] the neglecting of the limited number of data points and, therefore, of both
limitations in the resolution and the block length.
\end{description}
The point (a) is crucial, without this assumption (even with an
enormous data set) it is not possible to distinguish between 1) and
2).  One has to say that in the cases 4) and 5) at least in principle
it is possible to understand that the systems are ``trivial''
(i.e. not chaotic) but for this one has to use a very huge number of
data. For example Cohen et al. estimated that in order to distinguish
between 1) and 5) using realistic parameters of a
typical liquid, the number of data points required has to be at least
$\sim 10^{34}$. Let us remind that Gaspard et al. \cite{Gaspard98} used
$\sim 1.5 \cdot 10^5$ points!

It seems to us that the distinction between chaos and noise makes
sense only in very peculiar cases, e.g. very low dimensional systems.
Nevertheless even in such a case an entropic analysis can be unable to
recognize the ``true'' character of the system due to the lack of resolution. 
This is particularly evident in the comparison between the diffusive map (\ref{eq:3-1.1})
and the noisy map (\ref{eq:3-5}), if one only observes at scales
$\epsilon>\sigma$. 
According to the pragmatic proposal of classification discussed in
sect.~\ref{sec:3} one has that for $\sigma \leq \epsilon \leq 1$ both
the system (\ref{eq:3-1.1}) and (\ref{eq:3-5}), in spite of their
``true'' character can be classified as chaotic, while for $\epsilon
\geq 1$ both can be considered as stochastic.  

In this respect, the problem of the lack of resolution is even more
severe for high entropic systems. One can roughly estimate the
critical $\epsilon$ ($\epsilon_u$), below which the saturation can be
observed, to be $\epsilon_u \propto \exp(-h_{KS})$. Indeed
$\exp(h_{KS})$ estimates the number of symbols, i.e. cells of the
partition required to reconstruct the dynamics.  Therefore, in the
Brownian motion studied by Gaspard et al. \cite{Gaspard98}, where the
$KS$-entropy is expected to be proportional to the number of molecules
present in the fluid, the possible small $\epsilon_u$ is pushed on
scales far from being reachable with the finest experimental
resolution available.

%%%%%%%%%%%%%%%%%%%%%%%%%%%%%%%%%%%%%%%%%%%%%%%%%%%%%%%%%%%%%%%%
\section{Discussions}
\label{sec:6}
%%%%%%%%%%%%%%%%%%%%%%%%%%%%%%%%%%%%%%%%%%%%%%%%%%%%%%%%%%%%%%%%

Here we  briefly review two examples,  studied in details in 
\cite{Olbrich98,Cencini99}, showing that  high-dimensional 
systems can display non-trivial behaviors at varying the resolution 
scales. We believe this discussion can be useful in furtherly clarify
the subtle aspects of the distinction between stochastic and deterministic 
behaviors in dynamical systems.

Olbrich et al. \cite{Olbrich98} analyzed an open flow system described by
unidirectionally coupled map lattice:
\begin{equation}
x_j(t+1)=(1-\sigma) f(x_{j+1}(t))+\sigma x_j(t)
\label{eq:dresden}
\end{equation}
where $j=1,\dots,N$ denotes the site of a lattice of size $N$, $t$ the
discrete time and $\sigma$ the coupling strength.  A detailed
numerical study (also supported by analytical arguments) of the
$\epsilon$-entropy $h(\epsilon)$ at different $\epsilon$, in the limit
of small coupling, gives the following scale-dependent scenario: for
$1 \geq \epsilon \geq \sigma$ there is a plateau
$h(\epsilon)=\lambda_s$ where $\lambda_s$ is the Lyapunov exponent of
the single map $x(t+1)=f(x(t))$. For $\sigma \geq \epsilon \geq
\sigma^2$ another plateau appears at $h(\epsilon) \approx 2
\lambda_s$, and so on for $\sigma^{n-1} \geq \epsilon \geq \sigma^{n}$
one has $h(\epsilon) \approx n \lambda_s$.  Similar results hold for
the correlation dimension which increases step by step as the
resolution increases, showing that the high-dimensionality of the
system becomes evident only as $\epsilon \to 0$.

Therefore one has a strong evidence that the dynamics at different
scales is basically ruled by a hierarchy of low-dimensional systems whose
"effective'' dimension $n_{eff}(\epsilon)$ increase as $\epsilon$
decreases:
\begin{equation}
n_{eff} (\epsilon) \sim \left[ {\ln(1/\epsilon) \over \ln(1/\sigma)}\right]\,,
\end{equation}
where $[\dots]$ indicates the integer part.  In addition, for a
resolution larger than $\epsilon$, it is possible to find a suitable
low-dimensional noisy system (depending on $\epsilon$) which is able
to mimic $x_1(t)$ given by eq. (\ref{eq:dresden}).  It is interesting
to note that, looking at $h(\epsilon)$ on
an extended range of values of $\epsilon$, one observes for $\epsilon
\geq \sigma^{N}$ 
\begin{equation}
 h(\epsilon) \sim \ln{1\over \epsilon} 
\end{equation}
i.e. the typical behavior of a stochastic process.  Of course for
$\epsilon \leq \sigma^{N}$ one has to realize that the system is
deterministic and $h(\epsilon)\approx N \lambda_s$.
Even if this study mainly concerns the small unidirectional coupling,
the diffusive and strong coupling case deserves further analysis, 
it represents a first step toward the understanding of the issue
of data analysis of high-dimensional systems. 

Let us now briefly discuss the issue of the macroscopic chaos,
i.e. the hydro-dynamical-like irregular behavior of some global
observable, with typical times much longer than the times related to
the evolution of the single (microscopic) elements composing a certain
system. This interesting kind of dynamical behavior has been studied
in some recent works \cite{Cencini99,Shibata98} for globally coupled
maps evolving according to the equation: \be x_i(t+1)= (1-\sigma)
f_a(x_i(t))+ {\sigma \over N} \sum_{j=1}^{N} f_a(x_j(t))
\label{eq:gcm}
\ee 
where $N$
is the total number of elements and $f_a$ is a nonlinear function
depending on a parameter $a$.
Cencini et al. \cite{Cencini99} (see also Shibata and Kaneko \cite{Shibata98}
for a related work) studied the behavior of a global variable
(i.e. the center of mass) using the FSLE analysis.
Their results can be summarized as follows:
\begin{itemize}
\item at small $\epsilon\,\,(\ll 1/\sqrt{N})$ one recovers the
``microscopic'' Lyapunov exponent, i.e. $\lambda(\epsilon)\approx
\lambda_{micro}$
\item at large $\epsilon\,\,(\gg 1/\sqrt{N})$ one observe another
plateau (corresponding to what we can call the ``macroscopic''
Lyapunov exponent) $\lambda(\epsilon) \approx \lambda_{macro}$ which
can be much smaller than the microscopic one.
\end{itemize}
The above results suggest that at a coarse-grained level,
i.e. $\epsilon \gg 1/\sqrt{N}$, the system can be described by an
``effective'' hydro-dynamical equation (which in some cases can be
low-dimensional), while the ``true'' high-dimensional character
appears only at very high resolution, i.e.
$$
\epsilon \leq \epsilon_c \approx O\left({1 \over \sqrt{N}}\right).
$$
The presence of two plateaux for the FSLE at different length scales
is present in generic systems with slow and fast dynamics
\cite{Boffetta98}. The
interesting fact is that in systems like (\ref{eq:gcm}) the two
temporal scales are generated by the dynamics itself.

Let us stress that the behavior $h(\epsilon)=const$ at small
$\epsilon$ and $h(\epsilon)$ decreasing for larger $\epsilon$ is not a
peculiarity of the diffusive map (\ref{eq:3-1.1}).  In typical
high-dimensional chaotic systems one has $h(\epsilon)= h_{KS}\sim
O(N)$ for $\epsilon \leq \epsilon_c$ (where $N$ is the number of
degrees of freedom and $\epsilon_c \to 0$ as $N\to \infty$) while for
$\epsilon\geq \epsilon_c$, $h(\epsilon)$ decreases (often with a power
law). From this point of view, the fact that in certain stochastic
processes $h(\epsilon) \sim \epsilon^{-\alpha}$ can be indeed
extremely useful for modeling such high-dimensional systems. As a
relevant example we just mention the fully developed turbulence which
is a very high-dimensional system whose non infinitesimal (the
so-called inertial range) properties can be successfully mimicked in
terms of a suitable stochastic process \cite{Biferale98}.

%%%%%%%%%%%%%%%%%%%%%%%%%%%%%%%%%%%%%%%%%%%%%%%%%%%%%%%%%%%%%%%%
\section{Conclusions}
\label{sec:7}
%%%%%%%%%%%%%%%%%%%%%%%%%%%%%%%%%%%%%%%%%%%%%%%%%%%%%%%%%%%%%%%%
We have shown how an entropic analysis at different resolution scales
(in terms of $\epsilon$-entropy and Finite Size Lyapunov Exponent) of
a given data record allows us for a classification of the stochastic
or chaotic character of a signal. \\ In practice, without any
reference to a particular model, one can define the notion of
deterministic or chaotic behavior of a system on a certain range of
scales. In our examples we show that, according to the pragmatic
classification proposed in Sect. III, one can consider (on a certain
resolution) a system as random or deterministic independently from its
``true'' nature.  At first glance this can appear disturbing, however,
if one adopts a non ``metaphysical'' point of view there is an
advantage in the freedom of modeling the behavior of the system at
least if one is interested on a certain (non infinitesimal)
coarse-graining property.

%%%%%%%%%%%%%%%%%%%%%%%%%%%%%%%%%%%%%%%%%%%%%%%%%%%%%%%%%%%%%%%%
\acknowledgments
%%%%%%%%%%%%%%%%%%%%%%%%%%%%%%%%%%%%%%%%%%%%%%%%%%%%%%%%%%%%%%%%
We thank R. Hegger and T. Schreiber for useful discussions and suggestions. 
M.C., M.F. and A.V. have been partially supported 
by INFM (PRA-TURBO) and by the European Network {\it Intermittency 
in Turbulent Systems} (contract number FMRX-CT98-0175).

%%%%%%%%%%%%%%%%%%%%%%%%%%%%%%%%%%%%%%%%%%%%%%%%%%%%%%%%%%%%%%%%
%%%%%%%%%%%%%		FIGURES			%%%%%%%%%%%%%%%%
%%%%%%%%%%%%%%%%%%%%%%%%%%%%%%%%%%%%%%%%%%%%%%%%%%%%%%%%%%%%%%%%

%%%%%%%%%%%%%%%%%%%%%%%%%%%%%%%%%%%%
%\newpage
%%%%%%%%%%%%%%%%%%%%%%%%%%%%%%%%%%%%
%%%%%%%%%%%%%%%%%%%%%%%%%%%%%%%%%%%%
%\newpage
%%%%%%%%%%%%%%%%%%%%%%%%%%%%%%%%%%%%
%%%%%%%%%%%%%%%%%%%%%%%%%%%%%%%%%%%%
%\newpage
%%%%%%%%%%%%%%%%%%%%%%%%%%%%%%%%%%%%
%%%%%%%%%%%%%%%%%%%%%%%%%%%%%%%%%%%%
%\newpage
%%%%%%%%%%%%%%%%%%%%%%%%%%%%%%%%%%%%
%%%%%%%%%%%%%%%%%%%%%%%%%%%%%%%%%%%%
%\newpage
%%%%%%%%%%%%%%%%%%%%%%%%%%%%%%%%%%%%
%%%%%%%%%%%%%%%%%%%%%%%%%%%%%%%%%%%%
%\newpage
%%%%%%%%%%%%%%%%%%%%%%%%%%%%%%%%%%%%
%%%%%%%%%%%%%%%%%%%%%%%%%%%%%%%%%%%%
%\newpage
%%%%%%%%%%%%%%%%%%%%%%%%%%%%%%%%%%%%
%%%%%%%%%%%%%%%%%%%%%%%%%%%%%%%%%%%%
%\newpage
%%%%%%%%%%%%%%%%%%%%%%%%%%%%%%%%%%%%


\begin{thebibliography}{99}

\bibitem{Nicolis84}
C. Nicolis and G. Nicolis, Nature {\bf 311},  529  (1984);
P. Grassberger, Nature {\bf 323},  609  (1986);
C. Nicolis and G. Nicolis, Nature {\bf 326},  523  (1987);
P. Grassberger, Nature {\bf 326},  524  (1987).

\bibitem{Osborne89a}
A.~R. Osborne and A. Provenzale, Physica D {\bf 35},  357  (1989).

\bibitem{Sugihara90a}
G. Sugihara and R. May, Nature {\bf 344},  734  (1990).

\bibitem{Casdagli91}
M. Casdagli, J. Roy. Statist. Soc. Ser. B {\bf 54},  303  (1991).

\bibitem{Kaplan92a}
D.~T. Kaplan and L. Glass, Phys. Rev. Lett. {\bf 68},  427  (1992).

\bibitem{Kubin95} G. Kubin,  in {\em Workshop on Nonlinear Signal and Image
Processing, Vol. 1}, IEEE (IEEE, Halkidiki, Greece, 1995), pp.\ 141--145.

\bibitem{Aurell96}
E. Aurell, G. Boffetta, A. Crisanti, G. Paladin and 
A. Vulpiani, Phys. Rev. Lett. {\bf 77},  1262  (1996); J. Phys. A {\bf 30},  1  (1997).

\bibitem{Kolmogorov56}
A. Kolmogorov, IRE Trans. Inf. Theory {\bf 1},  102  (1956).

\bibitem{Shannon49}
C. Shannon and W. Weaver, {\em The Mathematical Theory of Communication}
  (University of Illinois Press, Illinois, 1993).

\bibitem{Gaspard93c}
P. Gaspard and X.-J. Wang, Physics Reports {\bf 235},  291  (1993).

\bibitem{Benzi85}
R. Benzi, G. Paladin, G. Parisi, and A. Vulpiani, J. Phys. A {\bf 18},  2157
  (1985).

\bibitem{Grassberger89}
P. Grassberger, Phisica Scripta {\bf 40},  346  (1989).

\bibitem{Lorenz69}
E. Lorenz, Tellus {\bf 21},  3  (1969).

\bibitem{Takens80}
F. Takens,  in {\em Dynamical Systems and Turbulence (Warwick 1980)}, Vol.~898
  of {\em Lecture Notes in Mathematics}, edited by D.~A. Rand and L.-S. Young
  (Springer-Verlag, Berlin, 1980), pp.\ 366--381.

\bibitem{Sauer91a}
T. Sauer, J.~A. Yorke, and M. Casdagli, J. Stat. Phys. {\bf 65},  579  (1991).

\bibitem{Fraser89}
A.~M. Fraser, IEEE Trans. Information Theory {\bf 35},  245  (1989);
D. Prichard and J. Theiler, Physica D {\bf 84}, 476 (1995).

\bibitem{Cohen85}
A. Cohen and I. Procaccia, Phys. Rev. A {\bf 31},  1872  (1985).

\bibitem{Grassberger83c}
P. Grassberger and I. Procaccia, Phys. Rev. A {\bf 28}, 2591 (1983).


\bibitem{Takens98}
F. Takens and E. Verbitski, Nonlinearity {\bf 11},  771  (1998).

\bibitem{Kantz97c}
H. Kantz and T. Schreiber, {\em Nonlinear Time Series Analysis} (Cambridge
  Univ. Press, Cambridge, UK, 1997).

\bibitem{Ding92}
M. Ding, C. Grebogi, E. Ott, T. Sauer,  J.A. Yorke, 
Physica D {\bf 69},  404  (1992); Phys. Rev. Lett. {\bf 70},  3872  (1993).

\bibitem{Olbrich97}
E. Olbrich and H. Kantz, Phys. Lett. A {\bf 232},  63  (1997).

\bibitem{Boffetta98}
G. Boffetta, A. Crisanti,  F. Paparella,  A. Provenzale and A. Vulpiani, 
Physica D {\bf 116},  301  (1998);
G. Boffetta, P. Giuliani, G. Paladin, and A. Vulpiani, J. Atm. Sci. {\bf 55},
  3409  (1998).

\bibitem{Grassberger83d} P. Grassberger and I. Procaccia, 
Physica D {\bf 9},  189  (1983).


\bibitem{Theiler91}
J. Theiler, Phys. Lett. {\bf 155},  480  (1991).

\bibitem{Kaplan93c}
D.~T. Kaplan and L. Glass, Physica D {\bf 64},  431  (1993).

\bibitem{Schell82}
M. Schell, S. Fraser, and R. Kapral, Physical Review A {\bf 26},  504  (1982).

\bibitem{Feller}
W. Feller, {\em An Introduction to Probability theory and its Applications}
  (Wiley and Sons, New York, 1968).

\bibitem{Provenzale92}
A. Provenzale, L. Smith, R. Vio, and G. Murante, Physica D {\bf 58},  31
  (1992).

\bibitem{Mazur60}
P. Mazur and E. Montroll, J. Math. Phys. {\bf 1},  70  (1960);
R.~I. Cukier and P. Mazur, Physica {\bf 53},  157  (1971).

\bibitem{Gaspard98}
P. Gaspard, M. E. Briggs, M. K. Francis, J. V. Sengers,
R.W. Gammon, J.R. Dorfman and R. V. Calabrese, Nature {\bf 394},  865  (1998).

\bibitem{Cohen_comment}
C. Dettman, E. Cohen, and H. van Beijeren, Nature {\bf 401},  875  (1999).

\bibitem{Grassberger_comment}
P. Grassbeger and T. Schreiber, Nature {\bf 401},  875  (1999).

\bibitem{Olbrich98}
E. Olbrich, R. Hegger, and H. Kantz, Phys. Lett. A {\bf 244},  538  (1998).

\bibitem{Cencini99}
M. Cencini, M. Falcioni, D. Vergni, and A. Vulpiani, Physica D {\bf 130},  58
  (1999).

\bibitem{Shibata98}
T. Shibata and K. Kaneko, Phys. Rev. Lett. {\bf 81},  4116  (1998).

\bibitem{Biferale98}
L. Biferale, G. Boffetta, A. Celani,  A. Crisanti and  A. Vulpiani,
 Phys. Rev. E {\bf 57},  R6261 (1998).


 
\end{thebibliography}
\end{document}